\documentclass[12pt,a4paper]{article}
\usepackage{amsmath}
\usepackage{amssymb}
\usepackage[cp866]{inputenc}
\usepackage[russian,ukrainian]{babel}

\begin{document}

\begin{center}
\textbf{On the Charge Dependence of the Pion-Nucleon Coupling Constant and
Nucleon-Nucleon Low-Energy Scattering Parameters}{\tiny \bigskip }

\textbf{V. A. Babenko\footnote{%
E-mail: pet2@ukr.net} and N. M. Petrov}

\textit{Bogolyubov Institute for Theoretical Physics, NAS of Ukraine, Kiev}
\end{center}

\thispagestyle{empty}

\noindent The simple model describing charge independence and charge
symmetry breaking of the pion-nucleon coupling constant is proposed. The
model, which has simple physical meaning and foundation, suggests directly
proportional dependence of the pion-nucleon coupling constants on the masses
of interacting nucleons and pions. Charge dependence of the pion-nucleon
coupling constant and the $^{1}S_{0}$-state low-energy nucleon-nucleon ($NN$%
) scattering parameters are studied on the basis of the Yukawa meson theory.
The values of the $np$ and $nn$ interaction characteristics calculated by
the proposed model agree quite well with the experimental data.{\scriptsize %
\bigskip }

\noindent PACS numbers: 13.75.Cs, 13.75.Gx, 14.20.Dh, 14.40.Be, 25.40.Cm,
25.40.Dn, 03.70.+k{\scriptsize \bigskip }

1. The pion-nucleon coupling constants are fundamental characteristics of
strong nuclear interaction. There are two types of pion-nucleon constants
--- pseudovector $f_{N\pi }$ and pseudoscalar g$_{N\pi }$, connected by the
relation g$_{N\pi }=\left( 2M_{N}/m_{\pi ^{\pm }}\right) f_{N\pi }$, where $%
M_{N}$ and $m_{\pi }$ are the masses of the interacting nucleon and pion.
The pion-nucleon coupling constants play an important role in the researches
of $NN$ and $\pi N$ interactions, and therefore great attention is
constantly paid to their study and refinement of their values [1--19]. The
situation development history regarding pion-nucleon coupling constants can
be found in [1--3]. At the moment, there are no serious disagreements on the
value of the neutral-pion coupling constant g$_{pp\pi ^{0}}^{2}$
characterizing the proton-proton interaction. One of the last experimentally
determined values of the pseudoscalar neutral-pion constant g$_{pp\pi
^{0}}^{2}=13.52\left( 23\right) $ [4] is in full compliance with the
previously found values of g$_{pp\pi ^{0}}^{2}=13.55\left( 13\right) $ [5]
and g$_{pp\pi ^{0}}^{2}=13.61\left( 9\right) $ [6].

With regard to the charged-pion coupling constant g$_{c}^{2}$, there is no
such agreement. The well-known compilation of Dumbrajs et al. [7] gives the
value of g$_{c}^{2}=14.28\left( 18\right) $ obtained in [8, 9] from data on $%
\pi ^{\pm }p$ scattering. On the basis of an energy-dependent partial-wave
analysis (PWA) of data on $NN$ scattering, the Nijmegen group found the
value of g$_{c}^{2}=13.54\left( 5\right) $ [1, 10] for the charged-pion
constant. This result was nearly coincident with coupling constant g$_{pp\pi
^{0}}^{2}=13.55\left( 13\right) $ determined for neutral pions by the same
group in [5]. The values g$_{c}^{2}\sim 13.7\div 13.8$ for the charged-pion
constant, which are close to the constant g$_{pp\pi ^{0}}^{2}\sim 13.5\div
13.6$ for neutral pions, were also obtained on the basis of data on $\pi
^{\pm }p$\ interaction in some other studies [11--14]. At the same time, the
Uppsala Neutron Research Group obtained much larger values for the
charged-pion coupling constant, g$_{c}^{2}=14.62\left( 35\right) $ [15], g$%
_{c}^{2}=14.52\left( 26\right) $ [16], and g$_{c}^{2}=14.74\left( 33\right) $
[17], which significantly exceed the average value of the coupling constant
for neutral pions, g$_{pp\pi ^{0}}^{2}=13.6$ [1--3]. Thus, at the moment, an
extremely important and fundamental unsolved question is the question of the
possible charge dependence of the pion-nucleon coupling constant, or, in
other words, the problem regarding difference between the pion-nucleon
coupling constants for neutral and charged pions [1--19]. In connection with
this discussion, we would also like to recall that the term
\textquotedblleft charge independence\textquotedblright\ is, in fact,
synonym for the term \textquotedblleft isospin invariance\textquotedblright\
[1--3]. In the present work, which is a continuation of our previous papers
[18, 19], we study charge independence and charge symmetry breaking of
nuclear forces regarding the pion-nucleon coupling constant and the
low-energy $NN$ scattering parameters on the basis of the proposed simple
phenomenological model, which is based on the mass difference between the
charged and neutral pions, and in view also of the mass difference between
the neutron and the proton. The proposed model is compatible well with the
standard classical Yukawa meson model [20].

2. With taking into account the conservation of electric charge in the
pion-nucleon system, one has to differentiate in a general way four
different types of elementary pseudovector pion-nucleon coupling constants
[1, 3]%
\begin{equation}
f_{p\pi ^{0}\rightarrow p}~,~~~f_{n\pi ^{0}\rightarrow n}~,~~~f_{p\pi
^{-}\rightarrow n}~,~~~f_{n\pi ^{+}\rightarrow p}~,  \tag{1}
\end{equation}%
which correspond to four possible types of elementary vertices of
interaction $p\pi ^{0}\rightarrow p$, $n\pi ^{0}\rightarrow n$, $p\pi
^{-}\rightarrow n$, $n\pi ^{+}\rightarrow p$. Hereinafter we'll use the
notations $f_{p\pi ^{0}}$, $f_{n\pi ^{0}}$, $~f_{p\pi ^{-}}$, $f_{n\pi ^{+}}$%
, where the subscript corresponds to the input channel of the reaction $N\pi
\rightarrow N^{^{\prime }}$. Often, for the amplitudes (1), the notations $%
f_{p}$, $f_{n}$, $f_{-}$, $f_{+}$ respectively are also used.

Since the pion-nucleon coupling constant $f_{N\pi }$ is a measure of the
force applied to a nucleon with a mass $M_{N}$ ($N=p,n$) from the side of
the pion field with a non-zero mass of a quantum of a given field $m_{\pi }$
($\pi =\pi ^{0},\pi ^{+},\pi ^{-}$), it would be natural to assume that the
pion-nucleon constant is proportional to the product of the masses of the
nucleon and the pion%
\begin{equation}
f_{N\pi }=CM_{N}m_{\pi }~.  \tag{2}
\end{equation}%
With this in mind, for elementary pion-nucleon constants we have 
\begin{equation}
f_{p\pi ^{0}}=CM_{p}m_{\pi ^{0}}~,  \tag{3}
\end{equation}%
\begin{equation}
f_{n\pi ^{0}}=CM_{n}m_{\pi ^{0}}~,  \tag{4}
\end{equation}%
\begin{equation}
f_{p\pi ^{-}}=CM_{p}m_{\pi ^{-}}~,  \tag{5}
\end{equation}%
\begin{equation}
f_{n\pi ^{+}}=CM_{n}m_{\pi ^{+}}~.  \tag{6}
\end{equation}%
It should be particularly emphasized that the proposed relation (2) is
actually an elementary generalization of the simple relation $f_{\pi ^{\pm
}}/f_{\pi ^{0}}=m_{\pi ^{\pm }}/m_{\pi ^{0}}~$, which was written for the
case of exact fulfillment of charge symmetry ($f_{\pi ^{0}}^{2}\equiv
f_{pp\pi ^{0}}^{2}=f_{nn\pi ^{0}}^{2}=f_{0}^{2}$~, $f_{\pi ^{\pm
}}^{2}=f_{c}^{2}$) and has been considered earlier in one form or another in
a number of papers [2, 19, 21--23]. The relation (2) has a simple physical
justification and interpretation [19]. Namely, due to the fact that the
pion-nucleon constants $f_{N\pi }$ serve as a measure of the force effect of
pions and nucleons on each other, it is natural to assume that this effect
will be greater, the greater the masses of particles involved in the
interaction are. Further calculations and conclusions show that the proposed
relation (2) leads to a number of reasonable results and consequences, which
agree well with the experimental data in a number of cases. Thus, the
proposed hypothesis is generally self-consistent and consistent with the
experimental data and the standard classical meson model of Yukawa alike. On
the whole, it should be particularly emphasized that the hypothesis that
charge independence breaking of nuclear forces in the nucleon-nucleon system
has its origin mainly in the difference of the masses of charged and neutral
pions is a very long-standing assumption of nuclear physics with a fairly
rich history and rationale [2, 22, 23].

Important combinations of elementary constants (1) are pion-nucleon coupling
constants, which characterize the force of the nuclear interaction between
two nucleons and can be generally defined as follows [1, 3]%
\begin{equation}
f_{pp\pi ^{0}}^{2}=f_{p\pi ^{0}}\cdot f_{p\pi ^{0}}~,  \tag{7}
\end{equation}%
\begin{equation}
f_{nn\pi ^{0}}^{2}=f_{n\pi ^{0}}\cdot f_{n\pi ^{0}}~,  \tag{8}
\end{equation}%
\begin{equation}
f_{0}^{2}=f_{p\pi ^{0}}\cdot f_{n\pi ^{0}}~,  \tag{9}
\end{equation}%
\begin{equation}
f_{c}^{2}=f_{p\pi ^{-}}\cdot f_{n\pi ^{+}}~.  \tag{10}
\end{equation}%
At the moment, of all types of pion-nucleon constants, the proton-proton
coupling constant $f_{pp\pi ^{0}}^{2}$ [5] has been determined
experimentally in the most reliable and accurate way. So, there are no
serious disagreements on the value of this constant. And then, the
neutron-neutron constant $f_{nn\pi ^{0}}^{2}$, the neutral-pion constant $%
f_{0}^{2}$ and the charged-pion constant $f_{c}^{2}$, in accordance with
Eqs. (7)--(10) with regard also to Eqs. (3)--(6), are expressed through the
constant $f_{pp\pi ^{0}}^{2}$ within the framework of the proposed model in
the following way%
\begin{equation}
f_{nn\pi ^{0}}^{2}=\frac{M_{n}^{2}}{M_{p}^{2}}f_{pp\pi ^{0}}^{2}\text{~}, 
\tag{11}
\end{equation}%
\begin{equation}
f_{0}^{2}=\frac{M_{n}}{M_{p}}f_{pp\pi ^{0}}^{2}\text{~},  \tag{12}
\end{equation}%
\begin{equation}
f_{c}^{2}=\frac{M_{n}}{M_{p}}\frac{m_{\pi ^{\pm }}^{2}}{m_{\pi ^{0}}^{2}}%
f_{pp\pi ^{0}}^{2}\text{~}.  \tag{13}
\end{equation}

In accordance with Eqs. (11)--(13), by using the reliably established
experimental value of the proton-proton constant [5]%
\begin{equation}
f_{pp\pi ^{0}}^{2}=0.0749\left( 7\right)  \tag{14}
\end{equation}%
and experimental values of the masses of nucleons and pions [24], we obtain
the following values for the pseudovector pion-nucleon constants $f_{nn\pi
^{0}}^{2}$, $f_{0}^{2}$ and $f_{c}^{2}$:%
\begin{equation}
f_{nn\pi ^{0}}^{2}=0.0751\left( 7\right) \text{~},  \tag{15}
\end{equation}%
\begin{equation}
f_{0}^{2}=0.0750\left( 7\right) \text{~},  \tag{16}
\end{equation}%
\begin{equation}
f_{c}^{2}=0.0802\left( 7\right) \text{~}.  \tag{17}
\end{equation}

The pseudoscalar pion-nucleon coupling constant g$_{N\pi }$ and the
pseudovector coupling constant $f_{N\pi }$ are related by the well-known
equivalence relation [1--3]%
\begin{equation}
\text{g}_{N\pi \rightarrow N^{^{\prime }}}=\frac{M_{N}+M_{N^{^{\prime }}}}{%
m_{\pi ^{\pm }}}f_{N\pi \rightarrow N^{^{\prime }}}\text{~}.  \tag{18}
\end{equation}%
With taking into account Eq. (18) and using the pseudovector constants
(14)--(17), we obtain the following values for the pseudoscalar pion-nucleon
coupling constants 
\begin{equation}
\text{g}_{pp\pi ^{0}}^{2}=13.54(13)\text{~},  \tag{19}
\end{equation}%
\begin{equation}
\text{g}_{nn\pi ^{0}}^{2}=13.61(13)\text{~},  \tag{20}
\end{equation}%
\begin{equation}
\text{g}_{0}^{2}=13.58(13)\text{~},  \tag{21}
\end{equation}%
\begin{equation}
\text{g}_{c}^{2}=14.52(13)\text{~}.  \tag{22}
\end{equation}%
The value (22) of the pseudoscalar charged-pion coupling constant g$_{c}^{2}$%
, found in such a way within the framework of the proposed model, completely
coincides with the experimental value g$_{c}^{2}=14.52(26)$ obtained by the
Uppsala Neutron Research Group [16]. The value (22) also practically
coincides with the value g$_{c}^{2}=$g$_{\pi ^{\pm }}^{2}=14.53(25)$ we
obtained earlier [18, 19] in the Yukawa pion-nucleon model by using $pp$ and 
$np$ low-energy scattering parameters. The rather \textquotedblleft
large\textquotedblright\ value (22) of the constant g$_{c}^{2}$, obtained
for the considered model, is also in good agreement with a number of other
values of the charged-pion coupling constant obtained by processing the
experimental data on the $NN$ and $\pi N$ interaction [7--9, 15, 17, 25, 26].

At the same time, a number of other experimental determinations [6, 10--14,
27, 28] give significantly smaller values for the charged-pion constant g$%
_{c}^{2}$, which turn out to be close to the value (21) of the neutral-pion
constant g$_{0}^{2}=13.58(13)$, that indicates on the possible charge
independence of the pion-nucleon coupling constant. Thus, the issue of
charge dependence or charge independence of pion-nucleon coupling constants $%
f^{2}$ and g$^{2}$ remains still unsolved and requires further experimental
and theoretical researches [1--19, 26--31]. Nevertheless, the results of
this work and the results of our previous papers [18, 19], based on the
Yukawa meson theory, provide some kind of additional evidence of the charge
independence breaking for the pion-nucleon coupling constant in accordance
with modern concepts of the microscopic theory of strong interaction ---
quantum chromodynamics (QCD). Since, from the microscopic point of view of
QCD, the charge independence breaking of nuclear forces in general, and with
respect to the pion-nucleon coupling constant in particular, has its origin
in the existing differences in the masses and charges of $u$- and $d$%
-quarks, the elementary \textquotedblleft components\textquotedblright\ of
hadrons, then the pion-nucleon coupling constant according to QCD is
subjected to charge independence breaking due to necessity. For more details
on the state of the situation with pion-nucleon constants and their
determination, see the recently published review [3].

Even more subtle and not too researched is the issue of the possible charge
symmetry breaking of the pion-nucleon coupling constant, i.e. the issue of
the difference of pion-nucleon constants corresponding to the proton-proton
and neutron-neutron interaction. For the pion-nucleon constant corresponding
to the neutron-neutron interaction, the ratio of neutral-pion pseudoscalar
constants g$_{n\pi ^{0}}\equiv \hspace{1pt}$g$_{nn\pi ^{0}}$ and g$_{p\pi
^{0}}\equiv \hspace{1pt}$g$_{pp\pi ^{0}}$, corresponding to the $nn$ and $pp$
interaction, is often considered in the literature. Within the framework of
the proposed model, this ratio according to Eqs. (11), (18) is determined by
the formula:%
\begin{equation}
\frac{\text{g}_{n\pi ^{0}}}{\text{g}_{p\pi ^{0}}}=\frac{M_{n}^{2}}{M_{p}^{2}}%
=1.0028~.  \tag{23}
\end{equation}%
The value (23) obtained in such a way almost coincides with the value of the
ratio g$_{n\pi ^{0}}/$g$_{p\pi ^{0}}=1.0027$ of neutral-pion pseudoscalar
constants corresponding to the experimental values $f_{pp\pi
^{0}}^{2}=0.0751\left( 6\right) $ and $f_{0}^{2}=0.0752\left( 8\right) $ of
pseudovector constants found in [32] by Nijmegen group on the basis of the
PWA of $NN$ scattering data in the energy region $T_{lab}\leqslant 350~$MeV.
The value (23) of this ratio obtained in the model under consideration also
complies well with the value g$_{n\pi ^{0}}/$g$_{p\pi ^{0}}=1.0038$ obtained
in [33] on the basis of the Cloudy Bag Model (CBM) and with the value g$%
_{n\pi ^{0}}/$g$_{p\pi ^{0}}=1.0023$ found in [34] using the Feynman diagram
method. Thus, in general, the values of the neutron-neutron g$_{nn\pi
^{0}}^{2}$ and the charged-pion g$_{c}^{2}$ coupling constants obtained in
the proposed model by using the well-known $pp$ constant g$_{pp\pi ^{0}}^{2}$
agree well with a number of both experimentally determined and model values
of these quantities.

3. The low-energy scattering parameters of the nucleon-nucleon
effective-range theory, the scattering length and the effective range, are
the fundamental characteristics of the $NN$ interaction and the nuclear
forces in general [2, 22, 35--41]. The degree of violation of charge
independence and charge symmetry of nuclear forces [2, 22], as well as the
properties and characteristics of various $NN$ potentials and other physical
parameters and properties of the $NN$ system [38--40], are estimated by the
effect on the change in these parameters and their values. And at the same
time the $NN$ scattering length is usually the most sensitive and
characteristic parameter with respect to small changes in the $NN$ potential
or some other characteristics of the system. To calculate and estimate the
low-energy $NN$ scattering parameters for the proposed model, we consider
the description of the $NN$ interaction by the classical nucleon-nucleon
potential, which follows from the meson field theory, the Yukawa potential,
which contains the pion-nucleon coupling constant as the initial parameter.
For the $NN$ interaction in the spin-singlet $^{1}S_{0}$ state, the Yukawa
potential has a simple form [18--20]%
\begin{equation}
V_{YUK}\left( r\right) =-V_{0}\frac{e^{-\mu r}}{\mu r}~;~~\mu =\frac{m_{\pi
}c}{\hbar }~,~~V_{0}=m_{\pi }c^{2}f_{\pi }^{2}\text{~}.  \tag{24}
\end{equation}%
Two protons or two neutrons interact via the exchange of a neutral pion, and
in this case, the parameters of the Yukawa potential $\mu _{pp}$, $%
V_{0}^{pp} $ and $\mu _{nn}$, $V_{0}^{nn}$ in accordance with (24) are then
determined by the neutral-pion mass $m_{\pi ^{0}}$ and the coupling
constants $f_{pp\pi ^{0}}^{2}$ and $f_{nn\pi ^{0}}^{2}$. But in the case of
neutron-proton interaction, the exchange occurs via both neutral and charged
pions. And in the latter case, one should determine the parameters $\mu
_{np} $ and $V_{0}^{np}$ of the potential (24) by employing an averaged pion
mass value $\overline{m}_{\pi }$ and an averaged neutron-proton coupling
constant $f_{np\pi }^{2}$ [18, 19, 42]: 
\begin{equation}
\overline{m}_{\pi }\equiv \frac{1}{3}\left( m_{\pi ^{0}}+2m_{\pi ^{\pm
}}\right) ~,~~f_{np\pi }^{2}\equiv \frac{1}{3}\left(
f_{0}^{2}+2f_{c}^{2}\right) ~.  \tag{25}
\end{equation}

Further, we will use the well-known proton-proton low-energy scattering
parameters as the \textquotedblleft input\textquotedblright\ initial
parameters of the model, just as the proton-proton $\pi N$-constant $%
f_{pp\pi ^{0}}^{2}$ was previously used and specified. Namely, we evaluate
the Yukawa proton-proton potential parameters $\mu _{pp}$ and $V_{0}^{pp}$
by using the experimental proton-proton scattering length $a_{pp}$ and
effective range $r_{pp}$. But in doing this, corrections associated with
electromagnetic interaction should be removed from the real experimental
values of the nuclear-Coulomb low-energy $pp$ scattering parameters. After
the removal of these corrections, the values of the purely nuclear
scattering length $a_{pp}$ and effective range $r_{pp}$ for $pp$ scattering
turn out to be equal [22]%
\begin{equation}
a_{pp}^{\text{\textit{expt}}}=-17.3(4)\,\text{fm~},~~r_{pp}^{\text{\textit{%
expt}}}=2.85(4)\,\text{fm~}.  \tag{26}
\end{equation}%
By employing the variable-phase approach [43] and the values of the $pp$
scattering parameters (26), we obtain the following values for the
parameters of the Yukawa potential (24) in the case of $pp$-interaction%
\begin{equation}
\mu _{pp}=0.8392\,\text{fm}^{-1}\text{~},~~V_{0}^{pp}=44.8259\,\text{MeV~}. 
\tag{27}
\end{equation}%
Note that all further calculations of the low-energy $NN$ scattering
parameters for the Yukawa potential are also made on the basis of the
variable-phase approach [43].

As shown in our previous works on this issue [18, 19], the neutron-proton
parameters of the Yukawa potential (24) $\mu _{np}$ and $V_{0}^{np}$ are
related to analogous parameters of the proton-proton interaction $\mu _{pp}$
and $V_{0}^{pp}$ in the following way%
\begin{equation}
\mu _{np}=\frac{\overline{m}_{\pi }}{m_{\pi ^{0}}}\mu _{pp}\text{~}%
,~~V_{0}^{np}=\frac{\overline{m}_{\pi }}{m_{\pi ^{0}}}\frac{f_{np\pi }^{2}}{%
f_{pp\pi ^{0}}^{2}}V_{0}^{pp}\text{~}.  \tag{28}
\end{equation}%
Similarly, the neutron-neutron parameters of the potential (24) $\mu _{nn}$
and $V_{0}^{nn}$ are also related to the parameters of the proton-proton
interaction by the relations%
\begin{equation}
\mu _{nn}=\mu _{pp}\text{~},~~V_{0}^{nn}=\frac{f_{nn\pi ^{0}}^{2}}{f_{pp\pi
^{0}}^{2}}V_{0}^{pp}\text{~}.  \tag{29}
\end{equation}%
In accordance with Eqs. (28), (29), taking also into account (11)--(13),
(25), (26) and using the values of the Yukawa potential parameters for the $%
pp$ interaction (27), as well as the values of the nucleon and pion masses
[24], then we calculate the following values of the parameters $\mu $ and $%
V_{0}$ for the Yukawa potentials of $np$ and $nn$ interaction%
\begin{equation}
\mu _{np}=0.8583\,\text{fm}^{-1}\text{~},~~V_{0}^{np}=48.0246\,\text{MeV~}, 
\tag{30}
\end{equation}%
\begin{equation}
\mu _{nn}=0.8392\,\text{fm}^{-1}\text{~},~~V_{0}^{nn}=44.9496\,\text{MeV~}. 
\tag{31}
\end{equation}

Calculated further in such a way, by using the obtained parameters (30) of
the Yukawa potential (24), the values of the singlet $np$ scattering length $%
a_{np}$ and the effective range $r_{np}$ for the proposed model%
\begin{equation}
a_{np}=-23.4(4)\,\text{fm~},~~r_{np}=2.70(5)\,\text{fm}  \tag{32}
\end{equation}%
agree well with the experimental values [35, 36, 40, 41]%
\begin{equation}
a_{np}^{\text{\textit{expt}}}=-23.715(8)\,\text{fm~},~~r_{np}^{\text{\textit{%
expt}}}=2.71(7)\,\text{fm~}.  \tag{33}
\end{equation}

In a similar manner, by using the Yukawa potential parameters (31), we
obtain the following values for the neutron-neutron low-energy scattering
parameters $a_{nn}$ and $r_{nn}$%
\begin{equation}
a_{nn}=-18.2(4)\,\text{fm~},~~r_{nn}=2.84(5)\,\text{fm~}.  \tag{34}
\end{equation}%
As a result, and taking into account the inaccuracy measures, the values
(34) of the quantities $a_{nn}$ and $r_{nn}$, calculated within the
framework of the proposed model, are in good agreement with their
experimental values%
\begin{equation}
a_{nn}^{\text{\textit{expt}}}=-18.6(5)\,\text{fm~},~~r_{nn}^{\text{\textit{%
expt}}}=2.83(11)\,\text{fm~},  \tag{35}
\end{equation}%
found in [44] from the reaction $\pi ^{-}$+$d\rightarrow \gamma $+$2n$. The
found values of the neutron-neutron low-energy scattering parameters (34)
are also in very good agreement with the values $a_{nn}=-18.38(55)\,$fm, $%
r_{nn}=2.84(4)\,$fm that we have obtained based on the analysis of the
difference in the binding energies of the $^{3}$H and $^{3}$He mirror nuclei
[45, 46]. Thus, the value (34) of the neutron-neutron scattering length,
obtained for the model under consideration, is in good agreement with the
averaged experimental value of this quantity $a_{nn}^{\text{\textit{expt}}%
}\simeq -18.5(3)\,$fm [2, 22].

In order to verify the charge independence of nuclear forces, it is
necessary to put together and compare the neutron-proton, neutron-neutron
and proton-proton forces at low energies. Since the $^{1}S_{0}$ state of the 
$NN$ system features a virtual level with a nearly zero energy, then the $NN$
scattering length in this state is the most sensitive parameter in relation
to small variations of the $NN$ potential. For this reason, the difference
between the averaged value of the $pp$ and $nn$ scattering length and the $%
np $ scattering length is often used as a quantitative measure for the
charge independence breaking (CIB) of nuclear forces in the $NN$ system [22]%
\begin{equation}
\Delta a_{\text{CIB}}\equiv \frac{1}{2}\left( a_{pp}+a_{nn}\right) -a_{np}%
\text{~}.  \tag{36}
\end{equation}%
According to (26), (33), (35), the experimental value of this difference is%
\begin{equation}
\Delta a_{\text{CIB}}^{\text{\textit{expt}}}=5.8(3)\,\text{fm~},  \tag{37}
\end{equation}%
which amounts to $\sim $30\% in relative units. That this difference is
nonzero highly beyond the experimental uncertainty indicates that the
hypothesis of charge independence of nuclear forces is violated at low
energies [2, 22]. The experimental value of the $pp$ scattering length $%
a_{pp}$ (26) and the values of the $np$ scattering length $a_{np}$ (32) and $%
nn$ scattering length $a_{nn}$ (34), calculated in the proposed model, lead
to the following value of the quantity $\Delta a_{\text{CIB}}$ in this model%
\begin{equation}
\Delta a_{\text{CIB}}^{\text{\textit{theor}}}=5.7(4)\,\text{fm~}.  \tag{38}
\end{equation}%
The theoretical value $\Delta a_{\text{CIB}}$ (38) obtained in such a way
agrees very well with the experimental value (37). Hence, within the
framework of the discussed model, the charge independence breaking of
nuclear forces is fully explained by the mass difference between the charged
and neutral pions and by the mass difference, also present, between the
neutron and the proton, the latter giving some minor contribution. In
contrast with this, only $\sim $50\% of the experimental difference $\Delta
a_{\text{CIB}}^{\text{\textit{expt}}}$ (37) was explained by the mass
difference between the charged and neutral pions in earlier works --- see
[2, 22, 23, 47, 48].

4. Finally, to summarize, it should be emphasized that the pion-nucleon
coupling constants are of primary importance first of all for the low-energy
nuclear physics. The latter is caused mainly by the fact that the pions are
the lightest ones among the mesons and therefore the Yukawa's pion exchange
process determines the most long-range part of the nucleon-nucleon
interaction, namely the so-called one-pion-exchange tail. All in all,
considering the fact that the pion, predicted by Yukawa [20], is the most
important meson for our theoretical interpretation and understanding of the
nuclear forces, then the comprehension and exact quantitative explanation of
the coupling and interaction between the pion and the nucleon is crucial for
the development and construction of the strong nuclear interaction theory in
general [1--3, 22, 35--37, 49, 50]. It's actually also interesting to note
that different approaches and methods regarding quantum mechanical problem
for the Yukawa potential model are also under study in the contemporary
literature --- see, for example, recent paper [51] and references therein.

So, all told, on the basis of the standard classical Yukawa hypothesis [20]
that the $NN$ interaction at low energies is due to the exchange by
particles with nonzero mass --- pions, a physically justified model of
pion-nucleon interaction was proposed, wherein the pion-nucleon coupling
constant, characterizing the interaction force of the nucleon with the pion
field, is proportional to the product of the nucleon and pion masses: $%
f_{N\pi }=CM_{N}m_{\pi }$. The proposed phenomenological model of charge
independence and charge symmetry breaking of the pion-nucleon coupling
constant describes the difference between the four types of elementary
pseudovector pion-nucleon constants by simple splitting formulas (3)--(6)
proportionally to the masses of nucleons and pions participating in the
interaction. Therefore, in general, this model establishes a connection
between different types of pion-nucleon coupling constants. Physical
justification and interpretation of the model lies in the fact that the
pion-nucleon constants $f_{N\pi }$ serve as a measure of the force effect of
pions and nucleons on each other, and therefore it is natural to assume that
this effect will be greater, the greater are the masses of particles
involved in the interaction. Consequently, it follows that the charge
independence and charge symmetry breaking of the pion-nucleon constant in
the proposed model is directly related to the mass differences of the
interacting particles --- nucleons and pions.

Further, by making use of the reliably established experimental value of the
proton-proton coupling constant $f_{pp\pi ^{0}}^{2}$, the values of the
charged-pion $f_{c}^{2}=0.0802\left( 7\right) $, the neutral-pion $%
f_{0}^{2}=0.0750\left( 7\right) $ and the neutron-neutron $f_{nn\pi
^{0}}^{2}=0.0751\left( 7\right) $ pseudovector coupling constants were
calculated on the basis of the proposed model. The value of the pseudoscalar
charged-pion coupling constant g$_{c}^{2}=14.52(13)$, found within the
framework of the discussed model, completely coincides with the experimental
value of the Uppsala Neutron Research Group [16] and is in good agreement
with a number of other values of the charged-pion coupling constant,
obtained in different studies [7--9, 15--19, 25, 26]. The value for the
ratio of neutron g$_{nn\pi ^{0}}$ and proton g$_{pp\pi ^{0}}$ pseudoscalar
pion-nucleon coupling constants g$_{nn\pi ^{0}}/$g$_{pp\pi ^{0}}=1.0028$,
also obtained in this model, is in good agreement with a number of other
values of this quantity found in various models [32--34]. The values of the
singlet scattering lengths and effective ranges $a_{np}=-23.4(4)\,$fm, $%
r_{np}=2.70(5)\,$fm for $np$-scattering and $a_{nn}=-18.2(4)\,$fm, $%
r_{nn}=2.84(5)\,$fm for $nn$-scattering, calculated within the framework of
the proposed model by making use of the well-known experimental low-energy $%
pp$ scattering parameters, agree well with the experimental values of these
quantities with taking into account the inaccuracy measures of the latters.
Thus, the charge independence breaking of nuclear forces is almost
completely explained by the mass difference between the charged and neutral
pions and by the mass difference between the neutron and the proton for the
model under consideration. As we see, the calculations carried out and the
conclusions obtained on their basis show that, in general, the proposed
model leads to a number of reasonable results and consequences that are in
good agreement with the experimental data.

\begin{center}
{\normalsize REFERENCES}
\end{center}

\begin{enumerate}
\item J. J. de Swart, M. C. M. Rentmeester, and R. G. E. Timmermans,
nucl-th/9802084.

\item R. Machleidt and I. Slaus, J. Phys. G \textbf{27}, R69 (2001).

\item E. Matsinos, arXiv:1901.01204 [nucl-th].

\item V. Limkaisang, K. Harada, J. Nagata, \textit{et al.}, Prog. Theor.
Phys. \textbf{105}, 233 (2001).

\item J. R. Bergervoet \textit{et al.}, Phys. Rev. C \textbf{41}, 1435
(1990).

\item R. A. Arndt, I. I. Strakovsky, and R. L. Workman, Phys. Rev. C \textbf{%
52}, 2246 (1995).

\item O. Dumbrajs, R. Koch, H. Pilkuhn, \textit{et al.}, Nucl. Phys. B 
\textbf{216}, 277 (1983).

\item D. V. Bugg, A. A. Carter, and J. R. Carter, Phys. Lett. B \textbf{44},
278 (1973).

\item R. Koch and E. Pietarinen, Nucl. Phys. A \textbf{336}, 331 (1980).

\item V. Stoks, R. Timmermans, and J. J. de Swart, Phys. Rev. C \textbf{47},
512 (1993).

\item R. A. Arndt, W. J. Briscoe, I. I. Strakovsky, \textit{et al.}, Phys.
Rev. C \textbf{69}, 035213 (2004).

\item R. A. Arndt, W. J. Briscoe, I. I. Strakovsky, \textit{et al.}, Phys.
Rev. C \textbf{74}, 045205 (2006).

\item D. V. Bugg, Eur. Phys. J. C \textbf{33}, 505 (2004).

\item V. Baru, C. Hanhart, M. Hoferichter, \textit{et al.}, Nucl. Phys. A 
\textbf{872}, 69 (2011).

\item T. E. O. Ericson, B. Loiseau, J. Nilsson, \textit{et al.}, Phys. Rev.
Lett. \textbf{75}, 1046 (1995).

\item J. Rahm, J. Blomgren, H. Cond\'{e}, \textit{et al.}, Phys. Rev. C 
\textbf{57}, 1077 (1998).

\item J. Rahm, J. Blomgren, H. Cond\'{e}, \textit{et al.}, Phys. Rev. C 
\textbf{63}, 044001 (2001).

\item V. A. Babenko and N. M. Petrov, Phys. At. Nucl. \textbf{79}, 67 (2016).

\item V. A. Babenko and N. M. Petrov, Phys. Part. Nucl. Lett. \textbf{14},
58 (2017).

\item H. Yukawa, Proc. Phys. Math. Soc. Jap. \textbf{17}, 48 (1935).

\item H. H\"{o}gassen and D. O. Riska, Report HU-TFT-88-48, Univ. of
Helsinki (Helsinki, 1988).

\item G. A. Miller, B. M. K. Nefkens, and I. \v{S}laus, Phys. Rep. \textbf{%
194}, 1 (1990).

\item R. Machleidt and M. K. Banerjee, Few-Body Syst. \textbf{28}, 139
(2000).

\item K. A. Olive \textit{et al.} (Particle Data Group), Chin. Phys. C 
\textbf{38}, 090001 (2014).

\item J. Hamilton and W. S. Woolcock, Rev. Mod. Phys. \textbf{35}, 737
(1963).

\item V. G. J. Stoks \textit{et al.}, Phys. Rev. Lett. \textbf{61}, 1702
(1988).

\item E. R. Arriola, J. E. Amaro, and R. N. Perez, Mod. Phys. Lett. A 
\textbf{31}, 1630027 (2016).

\item R. N. P\'{e}rez, J. E. Amaro, and E. R. Arriola, Phys. Rev. C \textbf{%
95}, 064001 (2017).

\item J. M. Alarcon, J. Martin Camalich, and J. A. Oller, Ann. Phys. \textbf{%
336}, 413 (2013).

\item E. Matsinos and G. Rasche, Int. J. Mod. Phys. A \textbf{28}, 1350039
(2013).

\item E. Matsinos and G. Rasche, Int. J. Mod. Phys. E \textbf{26}, 1750002
(2017).

\item R. A. M. Klomp, V. G. J. Stoks, and J. J. de Swart, Phys. Rev. C 
\textbf{44}, R1258 (1991).

\item A. W. Thomas, P. Bickerstaff, and A. Gersten, Phys. Rev. D \textbf{24}%
, 2539(R) (1981).

\item L. K. Morrison, Ann. Phys. \textbf{50}, 6 (1968).

\item W. O. Lock and D. F. Measday, \textit{Intermediate Energy Nuclear
Physics} (Methuen, London) 1970.

\item A. G. Sitenko and V. K. Tartakovskii, \textit{Lectures on the Theory
of the Nucleus} (Pergamon, Oxford) 1975.

\item G. E. Brown and A. D. Jackson, \textit{The Nucleon-Nucleon Interaction}
(North-Holland Pub., Amsterdam) 1976.

\item V. A. Babenko and N. M. Petrov, Phys. At. Nucl. \textbf{66}, 1319
(2003).

\item V. A. Babenko and N. M. Petrov, Phys. At. Nucl. \textbf{68}, 219
(2005).

\item V. A. Babenko and N. M. Petrov, Phys. At. Nucl. \textbf{70}, 669
(2007).

\item V. A. Babenko and N. M. Petrov, Phys. At. Nucl. \textbf{73}, 1499
(2010).

\item T. E. O. Ericson and M. Rosa-Clot, Nucl. Phys. A \textbf{405}, 497
(1983).

\item V. V. Babikov, Sov. Phys. Usp. \textbf{10}, 271 (1967).

\item B. Gabioud, J.-C. Alder, C. Joseph, \textit{et al.}, Phys. Lett. B 
\textbf{103}, 9 (1981).

\item V. A. Babenko and N. M. Petrov, Phys. At. Nucl. \textbf{77}, 549
(2014).

\item V. A. Babenko and N. M. Petrov, Phys. Part. Nucl. Lett. \textbf{12},
584 (2015).

\item A. Sugie, Prog. Theor. Phys. \textbf{11}, 333 (1954).

\item T. E. O. Ericson and G. A. Miller, Phys. Lett. B \textbf{132}, 32
(1983).

\item L. Hulth\'{e}n and M. Sugawara, \textit{The Two-Nucleon Problem}, in
Encyclopedia of Physics, Vol. 39 of \textit{Handbuch der Physik}, Ed. S. Fl%
\"{u}gge (Springer, Berlin) 1957.

\item T. Ericson and W. Weise, \textit{Pions and Nuclei} (Clarendon Press,
Oxford) 1988.

\item J. P. Edwards \textit{et al.}, Prog. Theor. Exp. Phys. \textbf{083A01}
(2017).
\end{enumerate}

\end{document}